\documentstyle[12pt]{article}

\begin{document}

\hfill{UTTG-18-93}

\vspace{24pt}

\begin{center}
{\large{\bf   Effective Action and Renormalization Group
Flow of Anisotropic Superconductors }}

\vspace{36pt}
Steven Weinberg\footnote{Research supported in part by the
Robert A. Welch
 Foundation and NSF Grant PHY 9009850. Internet address
weinberg@utaphy.ph.utexas.edu.}

\vspace{4pt}
Theory Group, Department of Physics, University of Texas\\
Austin, TX, 78712

\vspace{30pt}
{\bf Abstract}

\end{center}

\begin{minipage}{4.75in}

We calculate the effective action of a superconductor,
without assuming that either the electron-electron potential
or the Fermi surface obey rotational invariance.  This
approach leads to the same gap equation and equilibrium free
energy as more conventional methods.  The
results are used to obtain the Gell-Mann - Low
renormalization group
equations for the electron-electron potential.
\end{minipage}

\vfill

\baselineskip=24pt
\pagebreak
\setcounter{page}{1}
\noindent
{\bf 1. Introduction}

This paper aims to demonstrate the usefulness in the theory
of superconductivity of the effective action formalism of
quantum field theory\footnote{For a lucid review of the
effective action method and
references to earlier work, see S. Coleman, ref. [1].}.
Although the effective action
may be defined non-perturbatively, for our purposes it may
be taken simply as the sum of all connected
one-particle-irreducible vacuum diagrams in the presence of
any background field.  As we shall see, the usual
assumptions of the BCS superconductivity theory allow an
almost trivial calculation of the effective action, from
which one may obtain all of the properties of the
superconductor, including the gap field, free energy,
penetration depth, and so on.

As an additional application of this formalism, we shall
derive the renormalization group flow of the
electron-electron interaction, a subject that has
received increased
attention in the last few years[2].  The renormalization
group equation for the electron-electron potential is
obtained here from the condition that the effective action
expressed as a function of the gap field should be
renormalization group invariant.    From a practical point
of view, the most significant difference with most earlier
work
is that here we make no special assumptions about the form
of the Fermi surface or the rotational invariance of the
electron-electron potential.\footnote{After this work was
completed I learned that the
renormalization group equation for non-circular Fermi
surfaces in two dimensions had been derived by R. Shankar,
Rev. Mod. Phys., ref. [2].}  Just as in the
rotationally invariant case, it turns out that the
renormalization group equations may be expressed as
de-coupled equations for an infinite number of coupling
parameters.\\

\noindent
{\bf 2. Calculation of the Effective Action}

As usual in the Fermi theory of liquids, we assume that all
degrees of freedom may be ``integrated out'' except for
electrons (strictly speaking, quasiparticles with the
quantum numbers of electrons) in a narrow shell around a
Fermi surface, say of thickness $\mu$ in electron energy.
(In the approach to be used here, this condition on electron
wave numbers is implemented through the renormalization
procedure, and must be verified {\it a posteriori}.)  Also
as usual, we discard all interactions that become irrelevant
in the limit $\mu\rightarrow 0$.  This leaves just the
one-electron and two-electron terms in the Lagrangian:
\begin{eqnarray}
&& L= \sum_{s',s}\int d^3x
\;\psi^{\dagger}(\vec{x},s')\left[\left(-
i\frac{\partial}{\partial
t}+eA_0(\vec{x},t)\right)\delta_{s',s}+E_{s',s}(-
i\vec{\nabla}+e\vec{A}(\vec{x},t))\right]\,
\psi(\vec{x},s) \nonumber \\ &&
-\frac{1}{4}\sum_{s_1',s_2',s_1,s_2}
\int
d^3x_1'd^3x_2'd^3x_1d^3x_2\;V_{s_1',s_2',s_1,s_2}(\vec{x_1}'
,\vec{x_2}',\vec{x_1},\vec{x_2})\nonumber\\&&\quad\times\;
\psi^{\dagger}
(\vec{x_1'},s_1')\,\psi^{\dagger}(\vec{x_2'},s_2')\,
\psi(\vec{x_1},s_1)\,\psi(\vec{x_2},s_2)
\end{eqnarray}
where $A^\mu$ is an external electromagnetic vector
potential, and $s,s'$, etc. are spin indices summed over the
values $\pm\frac{1}{2}$.
[The coefficient of the first term is adjusted to be a
simple Kronecker delta by suitable definition of the
electron field operator $\psi(\vec{x},s)$.  The time
argument is suppressed in all field operators.]
  Also, as well known, the only diagrams that are not
suppressed by powers of $\mu$ as $\mu\rightarrow 0$ are
those whose structure constrains each interacting pair of
electrons to have opposite momenta (for very slowly varying
external fields), so that if one electron is on the Fermi
surface, then the other is also.  [The Fermi surface is
defined by the vanishing of any eigenvalue of the energy
matrix $E_{s',s}(\vec{p})$, which is understood to include a
chemical potential term.  Time-reversal invariance tells us
that if a momentum $\vec{p}$ is on the Fermi surface, then
so is $-\vec{p}$.]  In particular, the unsuppressed
vacuum diagrams are those that become disconnected if we
cut through an interaction vertex so as to separate the
incoming from the outgoing electrons.

Before setting out to calculate these diagrams, we will
introduce a pair field $\Psi$ by a familiar trick, generally
attributed to Hubbard and Stratonovich[3], which we extend
here to general potentials.  Add a term to the Lagrangian of
the form
\begin{eqnarray}
&&\Delta L = \frac{1}{4}\sum_{s_1',s_2',s_1,s_2}
\int
d^3x_1'd^3x_2'd^3x_1d^3x_2\;V_{s_1',s_2',s_1,s_2}(\vec{x_1}'
,\vec{x_2}',\vec{x_1},\vec{x_2})\nonumber\\&&\times\;
[\Psi^\dagger_{s_1',s_2'}(\vec{x_1'}, \vec{x_2'})-
\psi^{\dagger}
(\vec{x_1'},s_1')\,\psi^{\dagger}(\vec{x_2'},s_2')]\nonumber
\\&&\times\;[\Psi_{s_1,s_2}(\vec{x_1},\vec{x_2})-
\psi(\vec{x_1},s_1)\,\psi(\vec{x_2},s_2)]
\end{eqnarray}
and integrate over the pair field
$\Psi(\vec{x_1},\vec{x_2})$  as well as over the electron
field $\psi(\vec{x})$.  This clearly has no effect; the
action is quadratic in $\Psi$, so the path integral may be
evaluated by setting $\Psi$ equal to its equilibrium value
$$
\Psi_{s_1,s_2}(\vec{x_1},\vec{x_2})=\psi(\vec{x_1},s_1)\,
\psi(\vec{x_2},s_2)
$$
at which (2) vanishes.  Instead of integrating over the pair
field, we shall evaluate the effective action in the
presence of a background pair field, integrating over the
electron field.  The term (2) has been chosen to cancel the
term in the Lagrangian quartic in the electron field,
leaving just quadratic terms:

\begin{eqnarray}
&& L+\Delta L= \sum_{s',s}\int d^3x
\;\psi^{\dagger}(\vec{x},s')\left[\left(-
i\frac{\partial}{\partial
t}+eA_0(\vec{x},t)\right)\delta_{s',s}+E_{s',s}(-
i\vec{\nabla}+e\vec{A}(\vec{x},t))\right]\,
\psi(\vec{x},s) \nonumber \\ &&
-\frac{1}{4}\sum_{s_1',s_2',s_1,s_2}
\int
d^3x_1'd^3x_2'd^3x_1d^3x_2\;V_{s_1',s_2',s_1,s_2}(\vec{x_1}'
,\vec{x_2}',\vec{x_1},\vec{x_2})\nonumber\\&&\times\;
\left[\psi^{\dagger}
(\vec{x_1'},s_1')\,\psi^{\dagger}(\vec{x_2'},s_2')\,\Psi_{s_
1,s_2}(\vec{x_1}, \vec{x_2})+\psi
(\vec{x_1},s_1)\,\psi(\vec{x_2},s_2)\,\Psi^{\dagger}_{s_1',s
_2'}(\vec{x_1'}, \vec{x_2'})\right]\nonumber\\&&
+\frac{1}{4}\sum_{s_1',s_2',s_1,s_2}
\int
d^3x_1'd^3x_2'd^3x_1d^3x_2\;\Psi^\dagger_{s_1',s_2'}(\vec{x_
1'},\vec{x_2'})\;V_{s_1',s_2',s_1,s_2}(\vec{x_1}',\vec{x_2}'
,\vec{x_1},\vec{x_2})\;\Psi_{s_1,s_2}(\vec{x_1},\vec{x_2})
\nonumber\\&&{}
\end{eqnarray}

Now, the effective action in a background pair field is
given by the sum of all one-particle-irreducible vacuum
diagrams  --- that is, all vacuum diagrams that cannot
be disconnected by cutting through any one internal
line[1].  On the other hand, we have already mentioned that
in the
limit $\mu\rightarrow 0$, we are to keep only graphs that
can be disconnected by slicing through any electron-electron
interaction so as to separate incoming from outgoing
electrons.  In using (3), this means that we are to keep
only graphs that are disconnected by cutting through any
internal pair field line.  {\em Since we are to keep only
graphs that both are and are not disconnected by cutting
through any internal pair field line, we conclude that we
must keep only graphs that have no internal pair field lines
at all.}  There are just two such graphs; a tree graph
arising from the last term in (3), and a
one-electron-loop graph whose value is given by the
determinant of the ``matrix'' accompanying the terms in (3)
quadratic in the electron field:
\begin{eqnarray}
&&\Gamma[\Psi]=\frac{1}{4}\sum_{s_1',s_2',s_1,s_2}
\int dt \int
d^3x_1'd^3x_2'd^3x_1d^3x_2\;\Psi^\dagger_{s_1',s_2'}(\vec{x_
1'},\vec{x_2'})\;\nonumber\\&&\quad\quad\times\;
V_{s_1',s_2',s_1,s_2}(\vec{x_1}',\vec{x_2}',\vec{x_1},\vec{x
_2})\;\Psi_{s_1,s_2}(\vec{x_1},\vec{x_2})
\nonumber \\ && -\frac{i}{2}\,\ln\,{\rm
Det}\left[\begin{array}{cc}
{\cal A} & {\cal B} \\ {\cal B}^\dagger & -{\cal A}^T
\end{array}\right]+\frac{i}{2}\,\ln\,{\rm
Det}\left[\begin{array}{cc}
{\cal A} & 0 \\ 0 & -{\cal A}^T \end{array}\right]+\Gamma[0]
\end{eqnarray}
where ${\cal A}$ and ${\cal B}$ are the ``matrices'':
\begin{eqnarray}
&&{\cal A}_{s'\vec{x'}t',s\vec{x}t}=\left[\left(-
i\frac{\partial}{\partial
t}+eA_0(\vec{x},t)\right)\delta_{s',s}+E_{s',s}(-
i\vec{\nabla}+e\vec{A}(\vec{x},t))\right]\delta^3(\vec{x'}-
\vec{x})\delta(t'-t)\nonumber\\&&{}\\&&
{\cal B}_{s'\vec{x'}t',s\vec{x}t}=\Delta_{s's}(\vec{x'},
\vec{x})\delta(t'-t)
\end{eqnarray}
and $\Delta_{s's}(\vec{x'},\vec{x})$ is the ``gap'' field:
\begin{equation}
\Delta_{s's}(\vec{x'},\vec{x})\equiv -
\frac{1}{2}\sum_{\sigma',\sigma}\int d^3y
d^3y'\;V_{s's\sigma'
\sigma}(\vec{x'},\vec{x},\vec{y'},\vec{y})\Psi_{\sigma'
\sigma}(\vec{y'}\vec{y})
\end{equation}
The constant $\Gamma(0)$ represents the contribution of
electrons not near the Fermi surface, for which the
approximations made here are not applicable.

To avoid becoming lost in a cloud of indices, we now
specialize to the case of spin-independent forces and a
spin-singlet pair field, writing
\begin{eqnarray}
&&\Psi_{\frac{1}{2}\; -\frac{1}{2}}=-\Psi_{\frac{1}{2}
\;\frac{1}{2}}\equiv \Psi\quad\quad \Psi_{\frac{1}{2}
\;\frac{1}{2}}=\Psi_{-\frac{1}{2}\; -\frac{1}{2}}=0\\&&
V_{\frac{1}{2}\; -\frac{1}{2}\; \frac{1}{2}\; -\frac{1}{2}}-
V_{\frac{1}{2}\; -\frac{1}{2}\; -\frac{1}{2}\;
\;\frac{1}{2}}=-V_{-\frac{1}{2}\; \frac{1}{2}\; \frac{1}{2}
\;-\frac{1}{2}} + V_{\;-\frac{1}{2}\; \frac{1}{2} -
\;\frac{1}{2}\; \frac{1}{2}} \equiv 2V  \\&&
E_{s's}=E\,\delta_{s's}
\end{eqnarray}
(It is easy to extend the results here to systems with a
more general spin dependence, such as liquid He$^3$.)  It
follows from (8) and (9)  that
\begin{equation}
\Delta_{\frac{1}{2}\; -\frac{1}{2}}=-\Delta_{-\frac{1}{2}
\;\frac{1}{2}}\equiv \Delta\quad\quad \Delta_{\frac{1}{2}
\;\frac{1}{2}}=\Delta_{-\frac{1}{2}\; -\frac{1}{2}}=0
\end{equation}
where
\begin{equation}
\Delta(\vec{x'},\vec{x})\equiv -\int d^3y
d^3y'\;V(\vec{x'},\vec{x},\vec{y'},\vec{y})\,\Psi(\vec{y'}
\vec{y})
\end{equation}
The effective action is now
\begin{eqnarray}
&&\Gamma[\Psi]=\int dt \int
d^3x_1'd^3x_2'd^3x_1d^3x_2\;\Psi^\dagger(\vec{x_1'},\vec{x_2
'})\;V(\vec{x_1}',\vec{x_2}',\vec{x_1},\vec{x_2})\;\Psi(
\vec{x_1},\vec{x_2})
\nonumber \\ && -i\,\ln\,{\rm Det}\left[\begin{array}{cc}
{\cal A} & {\cal B} \\ {\cal B}^\dagger & -{\cal A}^T
\end{array}\right]+i\,\ln\,{\rm Det}\left[\begin{array}{cc}
{\cal A} & 0 \\ 0 & -{\cal A}^T \end{array}\right]+\Gamma[0]
\end{eqnarray}
\begin{eqnarray}
&&{\cal A}_{\vec{x'}t',\vec{x}t}=\left[-
i\frac{\partial}{\partial t}+e{\cal A}_0(\vec{x},t)+E(-
i\vec{\nabla}+e\vec{A}(\vec{x},t))\right]\delta^3(\vec{x'}-
\vec{x})\delta(t'-t)\nonumber\\&&{}\\&&
{\cal B}_{\vec{x'}t',\vec{x}t}=\Delta(\vec{x'},
\vec{x})\delta(t'-t)
\end{eqnarray}

\noindent
{\bf 3. Translationally Invariant Equilibrium}

First we consider the translationally invariant case, with
no external electromagnetic fields.  Then the pair and gap
fields can be put in the form
\begin{eqnarray}
&& \Psi(\vec{x'},\vec{x})=\int
d^3p\;e^{i\vec{p}\cdot(\vec{x'}-\vec{x})}\Psi(\vec{p})
\\&&
\Delta(\vec{x'},\vec{x})=(2\pi)^{-3}\int
d^3p\;e^{i\vec{p}\cdot(\vec{x'}-\vec{x})}\Delta(\vec{p})
\end{eqnarray}
The electron-electron potential appears here in the form
\begin{eqnarray}
&&\int d^3x_1 d^3x_2 d^3x_3 d^3x_4\;
e^{i\vec{p'}\cdot(\vec{x_1}-
\vec{x_2})}\,e^{i\vec{p}\cdot(\vec{x_3}-
\vec{x_4})}\,V(x_1,x_2,x_3,x_4)\nonumber\\&&\quad\quad\equiv
\Omega_4\,V(\vec{p'},
\vec{p})
\end{eqnarray}
where $\Omega_4$ is the spacetime volume
\begin{equation}
\Omega_4\equiv \int d^3x\int dt
\end{equation}
The effective potential $U[\Psi]$ is defined[1] as minus the
effective action per spacetime volume
\begin{eqnarray}
&&U[\Psi]\equiv -\Gamma[\Psi]/\Omega_4 \nonumber\\&&
=-\int d^3p\,
d^3p'\;\Psi^*(\vec{p'})V(\vec{p'},\vec{p})\Psi(\vec{p})
\nonumber\\&& +\frac{i}{(2\pi)^4}\int d\omega d^3p\;
\ln \left(1-\frac{|\Delta(\vec{p})|^2}{\omega^2-
E^2(\vec{p})+i\epsilon}\right)+U[0]
\end{eqnarray}
with
\begin{equation}
\Delta(\vec{p})=-\int d^3p'\;V({p},\vec{p'})\,\Psi(\vec{p'})
\end{equation}
(We are working here at zero temperature.  For non-zero
temperature, the integral over $\omega$ is of course
replaced with a sum over discrete Matsubara frequencies.)
Wick rotating, integrating over $\omega$, and expressing
$\Psi$ in terms of $\Delta$, this becomes
\begin{eqnarray}
&&U[\Delta]=-\int d^3p\, d^3p'\;\Delta^*(\vec{p'})V^{-
1}(\vec{p'},\vec{p})\Delta(\vec{p})
\nonumber\\&&
-\frac{1}{(2\pi)^3}\int d^3p
\Big[\sqrt{E^2(\vec{p})+|\Delta(\vec{p})|^2}\,-
\,E(\vec{p})\Big]\;+U[0]
\end{eqnarray}

We pause to note that the gap equation is obtained from the
condition that $\Delta(\vec{p})$ be at the stationary point
$\Delta_0(\vec{p})$ of $U[\Delta]$:
\begin{eqnarray*}
&&0=\left.\frac{\delta U[\Delta]}{\delta
\Delta^*(\vec{p})}\right|_{\Delta=\Delta_0}
\\&&= -\int d^3p'\;V^{-
1}(\vec{p},\vec{p'})\Delta_0(\vec{p'})
-
\frac{1}{2(2\pi)^3}\frac{\Delta_0(\vec{p})}{\sqrt{E^2(\vec{p
})+|\Delta_0(\vec{p})|^2}}
\end{eqnarray*}
or in a more familiar form
\begin{equation}
\Delta_0(\vec{p})=-\frac{1}{2(2\pi)^3}\int d^3p'
\frac{V(\vec{p},\vec{p'})\,\Delta_0(\vec{p'})}{\sqrt{E^2
(\vec{p'})+|\Delta_0(\vec{p'})|^2}}
\end{equation}
Also, the effective potential at this stationary ``point''
is the free energy density, which using (23) may be
expressed in terms of the gap field:
\begin{eqnarray}
&&F_{equilibrium}-F_{\Delta=0}=U[\Delta_0]-U[0]\nonumber\\&&
=\frac{1}{(2\pi)^3}\int d^3p
\left[\frac{|\Delta_0(\vec{p})|^2}{2\sqrt{E^2(\vec{p})+|
\Delta_0(\vec{p})|^2}}-
\sqrt{E^2(\vec{p})+|\Delta_0(\vec{p})|^2}
+E(\vec{p})\right]
\end{eqnarray}
This is the same as the result normally derived from the BCS
ground state wave function[4].  However, it should be noted
that (22) is {\em not} the same as the formula for the free
energy as a functional of $\Delta(\vec{p})$ calculated from
the BCS wave function\footnote{This may be obtained e. g.
from Eq. (4-64) of ref. [5], using the
relations  $\Psi_k=u_k\,v_k$ and $u_k^2+v_k^2=1$ to express
the coefficients $u_k$ and $v_k$ in terms of the pair field
$\Psi_k$, and then converting to the notation of the present
paper by replacing sums over the discrete index $k$ with
integrals over $\vec{p}$, and inserting appropriate factors
of $(2\pi)^3$.}:
\begin{equation}
F[\Psi]=+\int
d^3p\,d^3p'\;\Psi^*(\vec{p'})\,V(\vec{p'},\vec{p})\,\Psi
(\vec{p})\,+\,\frac{1}{(2\pi)^3}\int d^3p\,
E(\vec{p})\,\left[1-
\sqrt{1-4(2\pi)^6|\Psi(\vec{p})|^2}\right]\;.
\end{equation}
  This difference arises because the effective potential
does not have the interpretation of an energy density for a
composite field like $\Psi(\vec{p})$ or
$\Delta(\vec{p})$, except at the stationary point of the
effective potential[6].  Nevertheless, both (22) and (25)
yield the same gap equation, and the same value for the
equilibrium free energy.

So far, the limitation of electron momenta to a thin shell
around the Fermi surface has been left implicit.  To make
this limitation explicit, we will write the electron momenta
as
\begin{equation}
\vec{p}=\vec{k}+\hat{n}(\vec{k})\ell\quad\quad
d^3p=d^2k\,d\ell
\end{equation}
where $\vec{k}$ is on the Fermi surface (that is,
$E(\vec{k})=0$), and $\hat{n}(\vec{k})$ is the unit vector
normal to the Fermi surface at $\vec{k}$.  [Here $d^2k$
should be understood as $Jd\theta_1d\theta_2$, where
$\theta_1$ and $\theta_2$ are coordinates on the Fermi
surface, and $J$ is the Jacobian of the transformation from
$\vec{p}$ to $\ell,\theta_1$, and $\theta_2$.  For a
spherical Fermi surface, $\int d^2k$ just gives a factor
$4\pi k_F^2$.]  We will temporarily impose the condition
that electron momenta are close to the Fermi surface by
introducing a cut-off $\Lambda$ on $\ell$, chosen small
enough so that $\Lambda \ll |\vec{k}|$ for all $\vec{k}$ on
the Fermi surface, and so that $V(\vec{p'},\vec{p})$ and
$\Delta(\vec{p})$ change negligibly as $\ell$ and $\ell'$
vary from $0$ to $\Lambda$.
[This cut-off will eventually be obviated by the
introduction of a renormalized electron-electron potential.]
With this cut-off, we may approximate
\begin{equation}
E(\vec{p})=v_F(\vec{k})\ell
\end{equation}
where $v_F$ is the Fermi velocity:
\begin{equation}
v_F(\vec{k})=\left.\frac{\partial
E(\vec{k}+\hat{n}(\vec{k})\ell)}{\partial
\ell}\right|_{\ell=0}
\end{equation}
Eq. (22) may now be written
\begin{eqnarray}
&&U[\Delta]=-\Lambda^2\int d^2k
d^2k'\;\Delta^*(\vec{k'})V^{-
1}(\vec{k'},\vec{k})\Delta(\vec{k})
\nonumber\\&&
-\frac{1}{(2\pi)^3}\int_0^\Lambda d\ell\int d^2k\;
\Big[\sqrt{\ell^2v_F^2(\vec{k})+|\Delta(\vec{k})|^2}\,-
\,\ell v_F(\vec{k})\Big]\;+U[0]
\end{eqnarray}
so that $U[\Delta]$ is now defined as a functional of the
gap field on the Fermi surface only.
We define a renormalized electron-electron potential at a
renormalization scale $\mu$ by
\begin{equation}
V^{-1}_\mu(\vec{k'},\vec{k})\equiv-\left.\frac{\delta^2
U[\Delta]}{\delta\Delta^*(\vec{k'})\delta\Delta(\vec{k})}
\right|_{\Delta(\vec{k})=\Delta(\vec{k})^*=\mu}
\end{equation}
When we express the electron-electron potential $V$ in terms
of the renormalized potential $V_\mu$, Eq. (29) for the
effective potential becomes:
\begin{eqnarray}
&&U[\Delta]=-\int d^2k\, d^2k'\;\Delta^*(\vec{k'})V^{-
1}_\mu(\vec{k'},\vec{k})\Delta(\vec{k})
\nonumber\\&&
-\frac{1}{(2\pi)^3}\int_0^\Lambda d\ell\int d^2k\;
\left[\sqrt{\ell^2v_F^2(\vec{k})+|\Delta(\vec{k})|^2}\,-
\,\ell v_F(\vec{k})\right.
\nonumber\\&&-
\left.\frac{|\Delta(\vec{k})|^2}{2(\ell^2v_F^2(\vec{k})+\mu^
2)^{1/2}}+
\frac{\mu^2|\Delta(\vec{k})|^2}{4(\ell^2v_F^2(\vec{k})+\mu^2
)^{3/2}}
\right]\;+U[0]
\end{eqnarray}
The one-loop integral over $\ell$ now converges as
$\ell\rightarrow \infty$, so we may remove the cut-off, and
find
\begin{eqnarray}
&&U[\Delta]=-\int d^2k\, d^2k'\;\Delta^*(\vec{k'})V^{-
1}_\mu(\vec{k'},\vec{k})\Delta(\vec{k})
\nonumber\\&&
+\frac{1}{2(2\pi)^3}\int
d^2k\;\frac{|\Delta(\vec{k})|^2}{v_F(\vec{k})}\left[\ln
\left(\frac{|\Delta(\vec{k})|}{\mu}\right)-1\right]\;+\;U[0]
\end{eqnarray}
The corresponding gap equation is obtained from the
condition that this expression for the effective potential
be stationary:
\begin{equation}
\Delta_0(\vec{k})=\frac{1}{2(2\pi)^3}\int
d^2k'\;V_\mu(\vec{k},\vec{k'})\, v_F^{-
1}(\vec{k'})\Delta_0(\vec{k'})\left[\ln\left(\frac
{|\Delta_0(\vec{k'})|}{\mu}\right)-\frac{1}{2}\right]
\end{equation}
and the equilibrium free energy is
\begin{equation}
F_{equilibrium}=F_{\Delta=0}-\frac{1}{4(2\pi)^3}\int d^2k
\;\frac{|\Delta_0(\vec{k})|^2}{v_F(\vec{k})}
\end{equation}

These very simple results should not be taken entirely
literally, because the effective potential will always
contain ``irrelevant'' terms of higher order in $\Delta$
arising from degrees of freedom that have been integrated
out here.  In particular, note that according to (32) the
difference $U[\Delta]-U[0]$ vanishes for
$\Delta\rightarrow 0$, is negative for sufficiently small
$\Delta$, and goes to $+\infty$ for $\Delta\rightarrow
\infty$, so it must have a stationary point $\Delta_0\neq
0$,
at which the gap equation (33) is satisfied, for any
renormalized electron-electron potential, whether repulsive
or attractive.   But this solution should not be taken
seriously if it has
$\Delta_0$ so large that it is outside the range of validity
of these equations.  To be specific, it is easy to see from
(32) that if
$V_\mu(\vec{k'},{k})$ is positive (in the matrix sense) for
some $\mu$, then the minimum $\Delta_0(\vec{k})$ of the
effective potential
will have a scale $ ||\Delta_0|| \geq \sqrt{e}\mu$, where
the ``scale'' $||\Delta||$ of an arbitrary function $\Delta
(\vec{k})$ is defined by the condition
$$
\int d^2k\;\frac{|\Delta(\vec{k})|^2}{v_F(\vec{k})}\;\ln
\left( \frac{|\Delta(\vec{k})|}{||\Delta||}\right)\equiv 0.
$$
In particular, if $V_\mu(\vec{k'},{k})$ is a positive
``matrix'' for $\mu$ of the order of the Debye frequency,
then the solution of the gap equation (33) is physically
irrelevant.\\

\noindent
{\bf 4. Renormalization Group Flow}

In the Wilson approach to the renormalization group that is
most familiar in condensed matter physics, we would keep the
cut-off $\Lambda$ finite, and derive a renormalization group
equation for the $\Lambda$-dependence of the
unrenormalized electron-electron potential
$V(\vec{k'},\vec{k})$ from the condition that the effective
potential (29) should be $\Lambda$-independent.  This
approach would be possible here, but it would require the
introduction of ``irrelevant'' terms in
$U[\Delta]$ of higher order in $\Delta(\vec{k})$ to keep
$U[\Delta]$ $\Lambda$-independent for finite  $\Lambda$.  It
is much simpler to apply the older approach of Gell-Mann and
Low: the renormalization group equation in this approach is
the condition that $U[\Delta]$ is independent of the
arbitrary renormalization scale $\mu$:
\begin{equation}
\mu \, \frac{d}{d\mu} V^{-1}_\mu(\vec{k'},\vec{k})=
-\frac{\delta^2(\vec{k'}-\vec{k})}{2(2\pi)^3v_F(\vec{k})}
\end{equation}
or equivalently
\begin{equation}
\mu \, \frac{d}{d\mu} V_\mu(\vec{k'},\vec{k})=
\frac{1}{2(2\pi)^3}\int d^2 k''\;
V_\mu(\vec{k'},\vec{k''})\,v^{-
1}_F(\vec{k''})\,V_\mu(\vec{k''},\vec{k})
\end{equation}
This of course also implies that the solution of the gap
equation is independent of $\mu$.
For the special case of a spherical Fermi surface Eq. (36)
is a continuous version of the discrete renormalization
group equation of Benfatto and Gallavotti[2], provided we
identify their constant $\beta$ with $1/2(2\pi)^3v_F$.  Eq.
(36) also agrees with the results of Shankar[2], with
fairly obvious changes to convert his results from two to
three dimensions.

     These renormalization group equations can be separated
into equations for the eigenvalues $\lambda_n(\mu)$ of the
Hermitian kernel
\begin{equation}
K_\mu(\vec{k'},\vec{k})\equiv
\frac{1}{2(2\pi)^3}\;v_F^{-1/2}(\vec{k'})\,v_F^{-
1/2}(\vec{k})\,V_\mu(\vec{k'},\vec{k})\;.
\end{equation}
{}From either (35) or (36), we see that the eigenvectors of
$K$ are renormalization-group invariant, while the
eigenvalues are governed by the flow equations
\begin{equation}
\mu\frac{d}{d\mu}\lambda_n(\mu)=\lambda_n^2(\mu)\;.
\end{equation}
For a spherical Fermi surface these eigenvectors are just
the spherical harmonics $Y^m_\ell(\hat{k})$, but we see that
the decoupling of the eigenmodes of $K$ is in fact quite
general, not depending on rotational invariance.

A completely repulsive potential may be defined as one for
which all eigenvectors $\lambda_n$ are positive.  If this is
true at some starting scale $\mu_0$ (say, the Debye
frequency) then as
$\mu\rightarrow 0$ the eigenvalues stay positive and become
smaller, so nothing interesting happens.  This conclusion
may be altered by the inclusion of formally irrelevant
electron-electron couplings, as in the Kohn-Luttinger
effect, but is not directly affected by anisotropies.  A
completely or
partly attractive potential is one for which all or some of
the eigenvalues of $K$ are negative.   Any eigenvalue
that is negative at some $\mu_0$ becomes infinite at a
finite $\mu<\mu_0$, with the eigenvalue that is largest in
magnitude becoming infinite first.  As already mentioned,
this is the case where superconductivity actually occurs.

     It is not clear what is gained by this renormalization
group analysis.   Within Fermi liquid theory, the effective
potential arises solely
from tree and one-loop graphs, so we do not need to use the
renormalization group to make the sort of improvement in
perturbation theory familiar in quantum electrodynamics,
or to identify a weak coupling regime, as in quantum
chromodynamics or the theory of critical phenomena.  Of
course, one can go beyond tree and one-loop graphs, but this
would require that we take into account irrelevant terms in
the original Lagrangian, which would involve
additional microscopic information, not just renormalization
group equations.\\

\noindent
{\bf 5. Slowly Varying Electromagnetic and Goldstone Fields}

     Now let us return to the general case of a
superconductor in  a translationally-non-invariant external
electromagnetic field.  In the limit where this field has
very small frequencies and wave numbers (smaller than the
inverse correlation length), we can
integrate out all degrees of freedom except those that have
zero ``mass,'' in the sense that their frequency vanishes in
the limit of vanishing wave number.  For a superconductor
that is not close to a phase transition at which
superconductivity is lost, the only such ``massless'' degree
of freedom is the Goldstone mode, associated with the
spontaneous breakdown of electromagnetic gauge invariance
within the superconductor.  All of the classic exact
properties of superconductors (persistent currents, Meissner
effect, flux quantization, and Josephson frequency)
can be derived by considering only general properties of the
effective action for the Goldstone mode in the presence of
external electromagnetic fields[7].  But to derive values
for quantities like the penetration
depth in Type II superconductors, we need a detailed formula
for the effective action.

The effective action for a Goldstone mode is in general
obtained by setting all the fields in the effective action
equal to their equilibrium values, and then subjecting them
to a symmetry transformation with space-time dependent
parameters equal to the Goldstone fields.  In our case,
where the broken symmetry is electromagnetic gauge
invariance, this
means that we must make the replacement
\begin{equation}
\Delta(\vec{x'},\vec{x},t)\rightarrow e^{-
ie\phi(\vec{x',t})}
\Delta_0(\vec{x'}-\vec{x}) e^{-ie\phi(\vec{x},t)}
\end{equation}
where $\phi(\vec{x})$ is the Goldstone field, and
$\Delta_0(\vec{x'}-\vec{x})$ is the equilibrium gap field,
given by the Fourier transform of the solution of the gap
equation:
\begin{equation}
\Delta_0(\vec{x'}-\vec{x})= (2\pi)^{-3}\int
d^3p\;e^{i\vec{p}\cdot(\vec{x'}-\vec{x})}\Delta_0(\vec{p})
\end{equation}
Electromagnetic gauge invariance\footnote{The Lagrangian (1)
is gauge invariant for either a local electron-electron
potential, or an arbitrary electron-electron potential
with a suitable dependence on the external electromagnetic
fields.} then allows us to remove
the factors $ e^{-ie\phi(\vec{x'},t')}$ and $e^{-
ie\phi(\vec{x},t)}$ in Eq. (39), by replacing the
electromagnetic vector potential $A_\mu(\vec{x},t)$ with
$A_\mu(\vec{x},t) -\partial_\mu\phi(\vec{x},t)$.  In this
way, Eqs. (13)-(15) yield the effective action:
\begin{equation}
\Gamma[\phi,A]=\Gamma_{\Delta=0}[A]-i\,\ln\,{\rm
Det}\left[\begin{array}{cc}
{\cal A} & {\cal B} \\ {\cal B}^\dagger & -{\cal A}^T
\end{array}\right]+i\,\ln\,{\rm Det}\left[\begin{array}{cc}
{\cal A} & 0 \\ 0 & -{\cal A}^T \end{array}\right]
\end{equation}
where now
\begin{eqnarray}
&&{\cal A}_{\vec{x'}t',\vec{x}t}=\left[-
i\frac{\partial}{\partial t}+eA_0(\vec{x},t)-e\dot{\phi}
(\vec{x},t)+
E(-i\vec{\nabla}+e\vec{A}(\vec{x},t)-e\vec{\nabla}\phi
(\vec{x},t)) \right]\nonumber\\&&\quad\times\quad
\delta^3(\vec{x'}-\vec{x})\delta(t'-t)\\&&
{\cal B}_{\vec{x'}t',\vec{x}t}=\Delta_0(\vec{x'}-
\vec{x})\delta(t'-t)
\end{eqnarray}
Quantitative properties of the superconductor such as the
penetration depth can be read off from the expansion of Eq.
(41) in powers of $ A_0(\vec{x},t)-\dot{\phi} (\vec{x},t)$
and $\vec{A}(\vec{x},t)-\vec{\nabla}\phi (\vec{x},t)$.

\noindent
{\bf Acknowledgment}

I am grateful for helpful conversations with S. Coleman, D.
Fisher, B. Halperin, and J. Polchinski.  My work on this
subject was stimulated by reading the lectures of
Polchinski.

\pagebreak
\noindent
{\bf References}

\begin{enumerate}

\item
S. Coleman, {\it Aspects of
Symmetry} (Cambridge University Press, Cambridge, 1985), pp.
136-144; Also see S. Coleman and E. Weinberg, Phys. Rev. D7
(1973) 1888.

\item
G. Benfatto and G. Gallavotti, J. Stat. Phys. 59 (1990) 541;
Phys. Rev. 42 (1990) 9967; J. Feldman and E.
Trubowitz, Helv. Phys. Acta 63 (1990) 157; R.
Shankar, Physica A 177 (1991) 530; ``Renormalization
Group Approach to Interacting Fermions,'' Rev. Mod. Phys.,
to be published; J. Polchinski, ``Effective Field Theory and
the
Fermi Surface,'' Santa Barbara/Texas preprint NSF-ITP-92-
132/UTTG-20-92, to be published.

\item
R. L. Stratonovich, Sov. Phys. Dokl. 2 (1957) 416; J.
Hubbard, Phys. Rev. Lett. 3 (1959) 77.  Also see S.
Weinberg, in {\it Proceedings of the 1962 International
Conference on High Energy Physics} (CERN, Geneva, 1962), p.
683.

\item
J. Bardeen, L. N. Cooper, and J. R. Schrieffer, Phys. Rev.
108 (1957) 1175.

\item
P. De Gennes,
{\it Superconductivity of Metals and Alloys}, translated by
P. A. Pincus (Benjamin, New York, 1966), p. 110.

\item
S. Coleman, {\it Aspects of Symmetry} (Cambridge University
Press, Cambridge, 1985), footnote 3 on page 401.

\item
S. Weinberg, ``Superconductivity for Particular
Physicists,'' in {\em Fields, Symmetries, Strings ---
Festschrift for Yochiro Nambu}, Prog. Theor. Phys. Suppl.
No. 86 (1986) 43.
\end{enumerate}
\end{document}